\documentclass[12pt]{article}
\usepackage{epsfig,amsmath,amssymb,array,dcolumn,subfigure,rotating}
\def\ul#1#2{\textstyle{\frac#1#2}}

\def\Acal{\mbox{\boldmath $\mathbb A $}}
\def\Mcal{\mbox{\boldmath $\mathbb M $}}
\def\Dcal{\mbox{\boldmath $\mathbb D $}}
\def\Tcal{\mbox{\boldmath $\mathbb T $}}
\def\Pcal{\mbox{\boldmath $\mathbb P $}}

\begin{document}

\title{Van der Waals interactions across stratified media}
\author{R. Podgornik $\dagger$$\star$  and V.A. Parsegian $\dagger$ \\[3mm] 
$\dagger$ Laboratory of Physical and Structural Biology\\
NICHD, Bld. 9 Rm. 1E116 \\
National Institutes of Health, Bethesda, MD
20892-0924\\[3mm] 
$\star$ Faculty of Mathematics and Physics \\
University of Ljubljana,  Ljubljana, Slovenia \\ and \\Dept. of Theoretical Physics, J. Stefan Institute, Ljubljana, Slovenia}
\maketitle

\begin{abstract}
Working at the Lifshitz level, we investigate the van der Waals interactions across  a series of layers with a periodic motif. We derive the complete form of the van der Waals interaction as an explicit function of the number of periodic layers. We then compare our result with an approximation based on an anisotropic-continuum representation of the stratified medium. Satisfactory agreement between discrete-layer and continuum models is reached only for thicknesses of ten or more layers.

\end{abstract}

\section{Introduction}

Once one starts looking for them, multilayers are everywhere. (Bio)macromolecules such as phospholipid layers \cite{nagle1} are prone to create multilamellar assemblies either in solution or in apposition to supporting interfaces \cite{nagle2}. Layer upon layer of polymer multilayers, assembled either  electrostatically  \cite{decher1} or through interlayer hydrogen bonding \cite{sukhishvili} allow fabrication of multicomposite molecular assemblies of tailored architecture \cite{decher2} with important technological applications. Understanding molecular interactions in these systems is an important step in controlling the assembly process. Though these interactions are due to many different specific  processes, the van der Waals interactions are their common feature. Unless one is satisfied with the trivial, pairwise additive formulation of van der Waals interactions in multilayer geometries, their general and exact derivation on the Lifshitz level is complicated, abstruse and seldom attempted. 

Here we will set up a new approach to the van der Waals - Lifshitz interactions in multilayer geometries based on a recent reformulation of the Lifshitz theory \cite{rudi1} in terms of an algebra of 2x2 matrices that allows us to derive simple and transparent formulas for the van der Waals interactions across finely layered systems.  We use this reformulation to evaluate the interactions across a multilayer and then compare these with a continuum uniaxial-dielectric approximation based on  van der Waals interactions across anisotropic media. 

\section{Formalism}

\begin{figure}[h]
\begin{center}
\epsfxsize=8cm
\epsfig{file=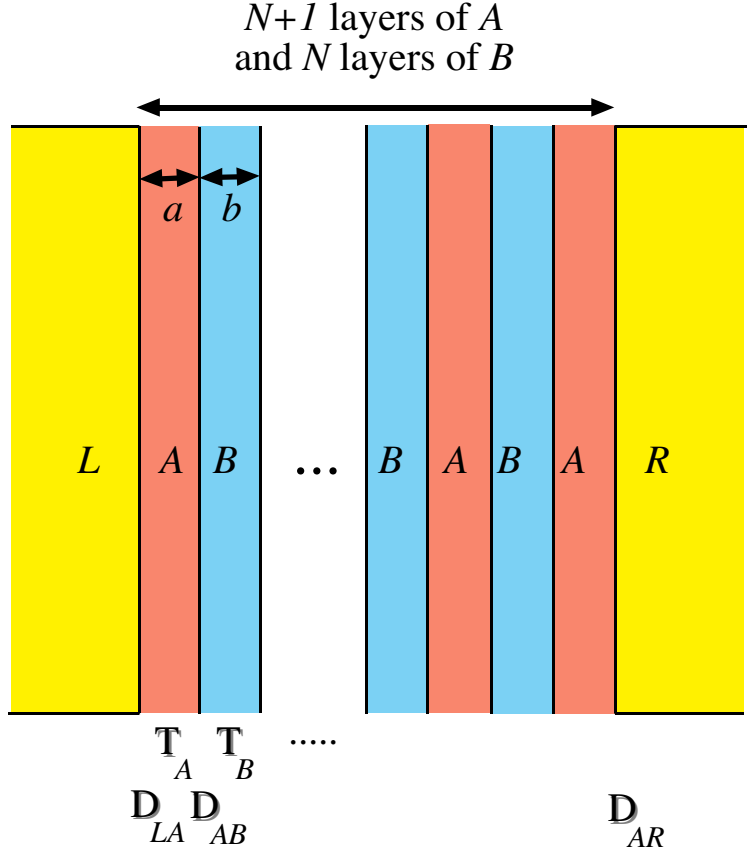, width=10cm}
\end{center}
\caption{Symmetric periodic array between a leftmost region $L$ and a rightmost region $R$. The periodic motif is the  $AB$ pair of layers repeated $N$ times between $L$ and $R$, symbolically  $L AB AB ... AB A R = L (AB)^N A R$. The thickness of layer $A$ (hydrocarbon) is $a$, and that of layer $B$ is $b$ (water). Matrices $ \Dcal_{LA},  \Dcal_{AB},  \Dcal_{AR}$ and $\mbox{\boldmath $\mathbb T $}_{A,B}$ are defined in the main text.}
\label{fig0}
\end{figure}

We consider a symmetric periodic array, Fig. \ref{fig0},  between a leftmost halfspace $L$ and a rightmost halfspace $R$. The periodic motif is the sequence of $AB$ pairs repeated $N$ times between $L$ and $R$, schematically $L AB AB ... AB A R$.
Recently we showed \cite{rudi1} that in the Lifshitz theory computation of the secular determinant of the electromagnetic field modes can be mapped onto an algebra of $2 \times 2$ matrices. The secular determinant in fact follows from the $11$ element of a transfer matrix that can be simply constructed from the interaction geometry as a product of discontinuity $\Dcal$ and propagator $\mathbb T$ matrices. In the case considered here, following the "mnemonic" introduced in \cite{rudi1}, the transfer matrix assumes the form
\begin{eqnarray}
	\Mcal = \Dcal_{RA}\times 	 \underbrace{ \underbrace{{\mbox{\boldmath$\mathbb T $}}_{A} \times \Dcal_{AB}\times \Tcal_{B} \times
 	\Dcal_{BA}}_{\Acal} \times 
	\dots \times \underbrace{{\mbox{\boldmath$\mathbb T $}}_{A} \times \Dcal_{AB}\times \Tcal_{B} \times
 	\Dcal_{BA}}_{\Acal}}_{N} \times  \Tcal_A \times {\mbox{\boldmath $\mathbb D 	$}}_{AL}.
\end{eqnarray}
In this notation the discontinuity and the propagator matrices become
\begin{equation}
\mbox{\boldmath $\mathbb D $}_{AB}=\left( 
\begin{array}{cc}
1 & - \overline{\Delta} \\ 
- \overline{\Delta} & 1
\end{array}
\right) = - \mbox{\boldmath $\mathbb D $}_{BA}, 
\end{equation}
and   
\begin{equation}
\mbox{\boldmath $\mathbb T $}_{A,B}=\left( 
\begin{array}{cc}
1 & 0 \\ 
0 & e^{-2\rho_{A,B}a,b}
\end{array}
\right),
\end{equation}
where $a$ and $b$ are the thicknesses of the $A$ and $B$ regions, and 
\begin{equation}
    \overline\Delta = \left(  \frac{\rho_{A}\epsilon_{B} -     \rho_{B}\epsilon_{A} }
     {\rho_{A}\epsilon_{B} +    \rho_{B}\epsilon_{A} } \right),
\end{equation}   
with  $\epsilon_{A}(\omega)$ and  $\epsilon_{B}(\omega)$ the frequency dependent dielectric functions of regions $A$ and $B$. Finally
\begin{equation}
\rho_{A,B} ^2 = Q^{2} - \frac{\epsilon_{A,B}\omega ^{2}}{c^{2}}. 
\end{equation}
We will assume that region $A$ corresponds to hydrocarbon, $\epsilon_{A}(\omega)$, and regions $L, B, R$ to water, $\epsilon_{B}(\omega)$, dielectric response. We use the standard \cite{ninham} forms for $\epsilon_{A}(\omega)$ and  $\epsilon_{B}(\omega)$ where the dielectric response of water is described with one MW relaxation frequency, five IR relaxation frequencies and six UV relaxation frequencies and that of the hydrocarbons with four UV relaxation frequencies.

The discontinuity matrix describes the propagation of the EM modes across the dielectric boundary and the propagator matrix their propagation inside a dielectrically homogeneous region. The above equations are strictly valid for the TM field modes. The result appropriate for the TE field modes is obtained analogously, {\sl via} a formal substitution 
\begin{equation}
	 \overline\Delta = \left(  \frac{\rho_{A}\epsilon_{B} -     \rho_{B}\epsilon_{A} }
     {\rho_{A}\epsilon_{B} +    \rho_{B}\epsilon_{A} } \right) \longrightarrow  \Delta = \left(  \frac{\rho_{A}\mu_{B} -     \rho_{B}\mu_{A} }
     {\rho_{A}\mu_{B} +    \rho_{B}\mu_{A} } \right),
\end{equation}
where $\mu_{A,B}$ are the magnetic permeabilities of the regions $A, B$.
\begin{figure}[h]
\begin{center}
\epsfxsize=8cm
\epsfig{file=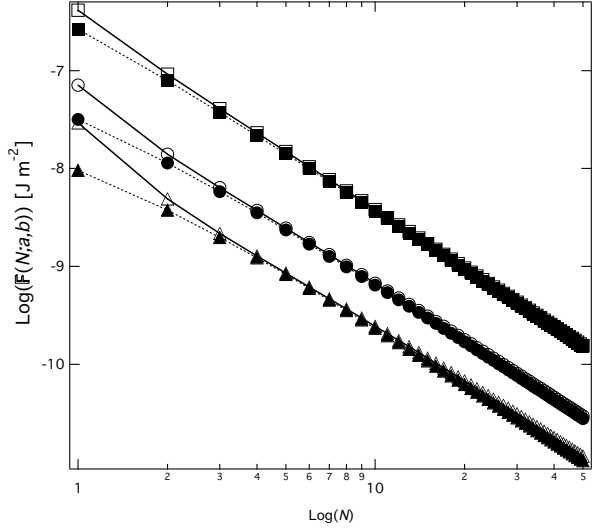, width=8cm}
\end{center}
\caption{Zero frequency  $n=0$ van der Waals interaction free energy across a multilayered slab of  layers $A$ and layers $B$; $a, b = 4 nm$ (circles); $a = 1nm$, $ b = 4 nm$ (triangles);  $a = 8nm$, $ b = 4 nm$ (squares). Region $A$ corresponds to lipid, regions $B,L, R$ to water with static dielectric functions \cite{ninham} $\epsilon_A = 2$ and $\epsilon_{B,L, R} = 80$. The result of the exact Eq. \ref{final} (bold curves) as well as approximate anisotropic continuum model Eq. \ref{aprox} (dashed curves) are presented. Clearly the exact and the approximate results differ only for small, $N \leq 10$, number of layers.}
\label{fig1}
\end{figure}
The transfer matrix can be written equivalently in the form
\begin{equation}
	\Mcal = \Dcal_{RA}\times 	\Acal^N \times {\mbox{\boldmath $
	\mathbb T $}}_{A} \times  {\mbox{\boldmath $\mathbb D 	$}}_{AL},
	\label{eki1}
\end{equation}	
where the matrix $\Acal$ can be obtained as
\begin{equation}
	\Acal = {\mbox{\boldmath$\mathbb T $}}_{A} \times \Dcal_{AB}\times \Tcal_{B} \times
 	\Dcal_{BA} =  \left( 
	\begin{array}{cc}
		1 - \overline{\Delta}^2 e^{- 2 \rho_{B} b} & \overline{\Delta}\left( 1 - e^{- 2 		\rho_{B} b} \right) \\ 
		- \overline{\Delta} e^{- 2 \rho_{A} a} \left( 1 - e^{- 2 		\rho_{B} b} \right) & e^{- 2 \rho_{A} a} \left( e^{- 2 \rho_{B} b} - \overline{\Delta}^2 		\right)
	\end{array}\right). 
\label{equ-a}
\end{equation}
The product $\Acal^N$ can be factored {\sl via} the Abel\' es formula for square matrices \cite{optics}.  This formula can be reproduced straightforwardly {\sl via} induction starting from the trivial $N = 2$ case  \cite{Abeles} so that
\begin{equation}
	\Acal^N = \frac{(\det \Acal)^{N/2}}{\sinh{\xi}}
	\left( 
	\begin{array}{cc}
		\sinh{N\xi} \frac{a_{11}}{\sqrt{\det \Acal}} - \sinh{(N-1)\xi} & \sinh{N\xi} 		\frac{a_{12}}{\sqrt{\det \Acal}} \\ 
		\sinh{N\xi} \frac{a_{12}}{\sqrt{\det \Acal}} & \sinh{N\xi} \frac{a_{11}}{\sqrt{\det \Acal}} - 		\sinh{(N-1)\xi}	
	\end{array}\right),
\end{equation}
\begin{figure}[h]
\begin{center}
\epsfxsize=8cm
\epsfig{file=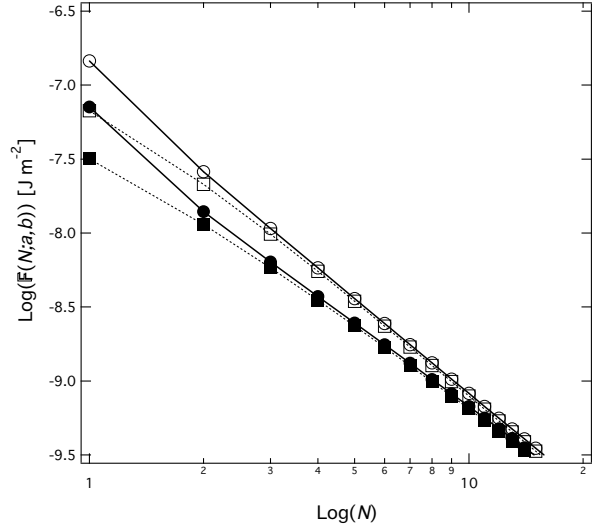, width=8cm}
\end{center}
\caption{Complete van der Waals interaction free energy Eq. \ref{final} across $N$ layers of hydrocarbon, $A$, and water, $B$, with thicknesses $a, b = 4 nm$.  The exact formula for $N$ layers, Eq. \ref{final}; open circles, bold lines for the complete $n$ sum, open squares, dashed line for the $n = 0$ term (as on Fig. \ref{fig1}). The approximate anisotropic continuum model, Eq. \ref{aprox}, with full circles, dashed lines for the complete $n$ sum and full squares, dashed line for the $n = 0$ term (same as on Fig. \ref{fig1}). Exact and approximate results again differ only for small, $N \leq 10$, number of layers.}
\label{fig3}
\end{figure}
where 
\begin{equation}
	\xi \equiv \log{\ul12 \frac{{\rm Tr}\Acal}{\sqrt{\det \Acal}} \left( 1 + \sqrt{1 - 4 \frac{\det \Acal}{({\rm Tr}\Acal)^2}} \right)}.
	\label{defxi}
\end{equation}
Here $e^{\xi}$ and $e^{-\xi}$ are the two eigenvalues of the unitary matrix $\Acal^* = \Acal/\sqrt{\det \Acal}$, {\sl viz.} for unitary matrices the product of the eigenvalues equals one. 
From the definition of $\xi$, Eq. \ref{defxi}, we derive explicitly
\begin{equation}
	\xi = (\rho_{A} a + \rho_{B} b) - \log{\frac{u}{1 + \sqrt{1 - \left( u e^{-(\rho_{A} a + \rho_{B} b)}\right)^2}}} = (\rho_{A} a + \rho_{B} b) - \log{ f(u, \rho_{A} a + \rho_{B} b  ) } ,
\end{equation}
where we defined
\begin{equation}
	u = \frac{2 ( 1-  \overline{\Delta}^2)}{1 -  \overline{\Delta}^2\left( e^{-\rho_{A} a} + e^{-\rho_{B} b} \right) + e^{- 2(\rho_{A} a + \rho_{B} b) }} \qquad {\rm and} \qquad f(u, \rho_{A} a + \rho_{B} b  ) = \frac{u}{1 + \sqrt{1 - \left( u e^{-(\rho_{A} a + \rho_{B} b)}\right)^2}}.
\end{equation}

The product $\Acal^N$ can be further decomposed as
\begin{equation}
	\Acal^N = \frac{(\det \Acal)^{N/2} ~ e^{(N-1) \xi}}{(1 - e^{- 2\xi})} ~\Pcal,
\end{equation}
where
\begin{equation}
	\Pcal(N) = \left( 
	\begin{array}{cc}
		p_{11}(N) &p_{12}(N) \\ 
		p_{21}(N) & p_{22}(N)	
	\end{array}\right) = 
	\left(1 - e^{- 2N \xi}\right) \frac{\Acal}{\sqrt{\det \Acal}} - e^{-\xi} \left( 1 - e^{- 2 (N-1) \xi} \right) \mathbb{I},
	\label{defp}
\end{equation}
with the unit matrix $\mathbb{I}$. 
\begin{figure}[h]
\begin{center}
\epsfxsize=8cm
\epsfig{file=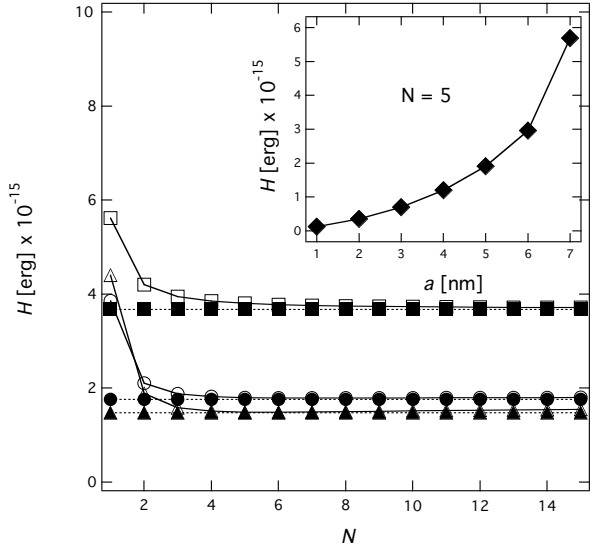, width=8cm}
\end{center}
\caption{The $n = 0$ Hamaker coefficient, Eq. \ref{hamaker}, of the van der Waals interaction across a multilayered slab of hydrocarbon layers $A$ and water layers $B$. Here  $a = 1nm$, $ b = 4 nm$ (triangles), $a, b = 4 nm$  (circles) and  $a = 8nm$, $ b = 4 nm$ (squares). Depending on the dielectric properties and the volume fractions of materials $A$ and $B$ the effect of the granularity persists for $N$ on the order of 10 layers. Inset: the variation of the  Hamaker coefficient at $N =5$ layers as a function of the thickness of the lipid layer, $a$ at $a + b = 8 nm$. Clearly the largest Hamaker coefficient is obtained for larger volume fractions of the lipid material.}
\label{fig2}
\end{figure}
In Lifshitz theory the fluctuation free energy is directly related to the $11$ element of the tranfer matrix $\Mcal$ \cite{rudi1}; here
\begin{equation}
	m_{11} =  \frac{(\det \Acal)^{N/2} ~ e^{(N-1) \xi}}{(1 - e^{- 2\xi})} \left( p_{11}(N) + p_{21}(N)   \overline{\Delta}_{RA} + e^{- 2 \rho_{A} a}\overline{\Delta}_{AL} ( p_{12}(N) + p_{22}(N) \overline{\Delta}_{RA})\right).
\end{equation}
The mode equation, or the secular determinant of the TM field  modes can be written
\begin{equation}
\label{ secular}
	m^{TM}_{11} = \mathcal{D}_{TM}(\omega,Q) = 0,
\end{equation}
with an analogous equation for TE field modes. The combined secular determinant thus equals the product $ \mathcal{D}(\omega,Q) = \mathcal{D}_{TM}(\omega,Q) \mathcal{D}_{TE}(\omega,Q) $. The free energy of the fluctuating EM modes in a system of $N$ (AB) layers can now be cast \cite{rudi1}  into a form containing the secular determinant of the TM and TE modes 
\begin{equation}
	\mathcal{F}(N; a, b) = {kT}{\sum_{\mathbf{Q}}\sum_{n=0}^{\infty }}^{\prime }\ln {\mathcal{D}(i\xi 	_{n},Q)} = {kT}{\sum_{\mathbf{Q}}\sum_{n=0}^{\infty }}^{\prime }\ln {m^{TM}_{11}(i\xi 	_{n},Q)} + {kT}{\sum_{\mathbf{Q}}\sum_{n=0}^{\infty }}^{\prime }\ln {m^{TE}_{11}(i\xi 	_{n},Q)},
\label{energy}
\end{equation}
where $\omega$ has been substituted by the imaginary Matsubara frequencies $i\xi_{n} = i 2\pi \frac{kT}{\hbar} n$ and the primed sum signifies that the $n = 0$ term is taken with the weight $1/2$. What we will investigate now is the dependence of the van der Waals interaction free energy, defined as the difference
\begin{equation}
\mathbb{F}(N; a, b) = \mathcal{F}(N; a, b) - \mathcal{F}(N \longrightarrow \infty; a, b)
\label{fren}
\end{equation}
on the number of (AB) layers, $N$. From the above definition of the interaction free energy we obtain the corresponding Hamaker coefficient $H(N; a, b)$ as
\begin{equation}
\mathbb{F}(N; a, b) = \frac{H(N; a, b)}{12 \pi (a+b)^2 N^2}.
\label{hamaker}
\end{equation}

\section{Result}

We first write the $11$ element for the TM modes in a form that makes explicit the dependence on $N$. We first notice that
\begin{equation}
	\log{m_{11}} = - \log{2 \sinh{\xi}} + N \log{ \sqrt{\det \Acal} ~e^{\xi}}  + \log{\left( p_{11}(N) + 	p_{21}(N)   \overline{\Delta}_{RA} + e^{- 2 \rho_{A} a}\overline{\Delta}_{AL} ( p_{12}(N) + p_{22}(N) 	\overline{\Delta}_{RA})\right) },
	\label{new.1.1}
\end{equation}
so that the corresponding free energy Eq. \ref{energy} has three terms. The first does not depend on $N$ and is thus not of immediate concern here. The second term, linear in $N$, represents the chemical potential or the free energy of adding a single combined layer to the system. As already formulated in \cite{adrian1} this chemical potential has the form
\begin{equation}
	\mu = \log{ \sqrt{\det \Acal} ~\lambda^{(+)}},
\end{equation}
where $\lambda^{(+)} = e^{\xi}$ is one of the eigenvalues of the matrix $\Acal^*$. However it is the last term of Eq. \ref{new.1.1} that describes the interactions in the multilayer composite. 
Using Eqs. \ref{energy} and \ref{fren} we end up with
\begin{equation}
	\mathbb{F}(N; a, b)  = kT~\sum_{\mathbf{Q}} {\sum_{n=0}^{\infty }}^{\prime } 
	\log{\left( \frac{p_{11}(N) + p_{21}(N)   \overline{\Delta}_{RA} + e^{- 2 \rho_{A} a}\overline{\Delta}_{AL} ( p_{12}(N) + p_{22}(N) \overline{\Delta}_{RA})}{p_{11}(\infty) + p_{21}(\infty)   \overline{\Delta}_{RA} + e^{- 2 \rho_{A} a}\overline{\Delta}_{AL} ( p_{12}(\infty) + p_{22}(\infty) \overline{\Delta}_{RA})}\right)},
\label{main.eq}
\end{equation} 
the main result of this paper. It gives the complete van der Waals interaction across the stratified medium $AB$ in a form with explicit dependence on the number of layers $N$. Not excessively complicated, it allows for straightforward numerical computations. 

Before analyzing Eq. \ref{main.eq} in detail, we verify that it has the correct limiting behavior. When $\epsilon_{A} = \epsilon_{B}$, so that $\overline{\Delta} = 0$ with $\rho_{A} = \rho_{B} = \rho$, if the stratified medium behaves homogeneously. In this case  
\begin{equation}
f(u, \rho_{A} a + \rho_{B} b  ) = 1 \qquad {\rm and} \qquad \xi = (\rho_{A} a + \rho_{B} b) = \rho (a + b).
\end{equation}
Furthermore in the same limit we obtain for the matrix $\Pcal$  
\begin{equation}
	\Pcal =  2 \sinh{(\rho_{A} a + \rho_{B} b)}\left( 
	\begin{array}{cc}
		1 & 0 \\ 
		0 & e^{- 2 N (\rho_{A} a + \rho_{B} b)}
	\end{array}\right).
\end{equation}
From these Eq. \ref{main.eq} becomes
\begin{equation}
	\mathbb{F}(N; a, b)  = kT~{\sum_{\mathbf{Q}}\sum_{n=0}^{\infty }}^{\prime } \log{\left( 1 + {\overline{\Delta}_{RA}\overline{\Delta}_{AL} e^{- 2 \rho ( a + N (a + b))}}\right)}.
\label{lif.eq}
\end{equation} 
Thus we recover the standard Lifshitz expression for the interaction of media $L$ and $R$ across the medium $A=B$ of thickness $ a + N (a + b)$. Obviously in the limit where the dielectric properties of $A$ and $B$ coincide, this is just the total thickness of the region $A$ . 

The general formula Eq. \ref{main.eq} can be rewritten 
\begin{equation}
	\mathbb{F}(N; a, b) =  \frac{kT}{2 \pi}~{\sum_{n=0}^{\infty }}^{\prime }\int_0^{\infty}\!\!\!\!\!Q dQ~\log{\left( \frac{p_{11}(N) + p_{21}(N)   \overline{\Delta}_{RA} + e^{- 2 \rho_{A} a}\overline{\Delta}_{AL} ( p_{12}(N) + p_{22}(N) \overline{\Delta}_{RA})}{p_{11}(\infty) + p_{21}(\infty)   \overline{\Delta}_{RA} + e^{- 2 \rho_{A} a}\overline{\Delta}_{AL} ( p_{12}(\infty) + p_{22}(\infty) \overline{\Delta}_{RA})}\right)}.
	\label{final}
\end{equation}
Its consequences can be fully appreciated only after a full numerical analysis. In Fig. \ref{fig1} we evaluate the $n=0$ term separately and then compare it to the full $n$ summation that includes retardation effects, Fig.  \ref{fig3}.

\section{Numerical computations and Conclusions}

For intuition about the general result Eq. \ref{final} we evaluate the van der Waals interactions across a slab of material of thickness $a + N(a + b)$, the continuum composite of layers $A$ and $B$ of Fig. \ref{fig0}. Because of its layered structure, we can associate with it a transverse dielectric function $\epsilon_{\perp} = \epsilon_{xx} = \epsilon_{yy}$ and a longitudinal dielectric  function  $\epsilon_{\parallel} = \epsilon_{zz}$. The continuum composite is thus a uni-axial dielectric with two distinct values of the dielectric response parallel and perpendicular to the layer normal. In terms of the dielectric functions $\epsilon_{A}$ and $\epsilon_{B}$ by analogy to capacitors in series and in parallel,
\begin{equation}
\epsilon_{\perp} = \frac{1}{a+b} \left(  a \epsilon_{A} + b \epsilon_{B} \right) \qquad {\rm and} \qquad \frac{1}{\epsilon_{\parallel}} = \frac{1}{a+b} \left(  \frac{a}{\epsilon_{A}} +  \frac{b}{\epsilon_{B}} \right).
\end{equation}
The longitudinal and the transverse dielectric response of a layered continuum composite thus depend on the volume fractions $a/(a+b)$ and $b/(a+b)$ of the materials $A$ and $B$ in the system. 

We use the result of \v Sarlah and \v Zumer \cite{sarlah} to derive the van der Waals interactions in a non-isotropic homogeneous uniaxial slab composed of periodic layers. If we associate the index $\overline{A}$ with the continuum uniaxial composite, we have for the TM modes 
\begin{equation}
\rho^2_{\overline{A}} = \frac{\epsilon_{\perp}}{\epsilon_{\parallel}} \left(  Q^2 - \frac{\epsilon_{\parallel} \omega^2}{c^2}\right).
\end{equation} 
The corresponding values for 
\begin{equation}
	 \overline\Delta_{R\overline{A}} = \left(  \frac{\rho_{\overline{A}}\epsilon_{A} -     \rho_{A}\epsilon_{\perp} }
     {\rho_{\overline{A}}\epsilon_{A} +    \rho_{A}\epsilon_{\perp} } \right) =  - \overline\Delta_{\overline{A}L},
\end{equation}
and {\sl mutatis mutandis} \cite{sarlah} for the TE field modes.Thus we obtain the interaction free energy as
\begin{equation}
	\mathbb{F}(N; a, b)  = \frac{kT}{2\pi} {\sum_{n=0}^{\infty }}^{\prime }\int_0^{\infty}\!\!\!\!\!Q dQ~\log{\left( 1 + {\overline{\Delta}_{R{\overline{A}}}\overline{\Delta}_{{\overline{A}}L} e^{- 2 \rho_{\overline{A}} (a +N (a + b))}}\right)},
\label{aprox}
\end{equation} 
plus an analogous term for the TE modes. This is the expression that we will compare with our exact formula Eq. \ref{final}.

Fig. \ref{fig1} gives the dependence of the zero order term, $n = 0$, interaction free energy $\mathbb{F}(N; a, b)$, in its exact form Eq. \ref{final} as well as its continuum approximate form Eq. \ref{aprox}, on the number of layers $N$, where we have taken $A$ to correspond to hydrocarbon and $B, L, R$ to correspond to water. The cases for $b = 4 nm$ and $a = 1,4,8 nm$ are shown. It is clear from Fig. \ref{fig1} that the effect of discreteness of layers is effectively gone after about 10 layers in the stack. Also obviously the interaction free energy is larger the larger the volume fraction of the lipid in the system.

The same is true also for the complete form, Eq. \ref{final}, with the explicit summation over $n$, Fig. \ref{fig3}. The higher order terms in the $n$-summation are obviously important only for small $N$. In both cases, $n=0$ as well as full summation over $n$, the way the continuum limit is approached  obviously depends on the characteristics of the layers, especially on their respective thicknesses. As is well known \cite{ninham} for the hydrocarbon-water system the $n = 0$ term gives the dominant contribution to the interaction energy and assures that contrary to the case of other materials the van der Waals interactionshave a non-retarded form even at very large separations.

The $n = 0$ Hamaker coefficient, Fig. \ref{fig2} and Eq. \ref{hamaker}, reinforces conclusions based on the evaluation of the interaction free energy. Again  the continuum limit of the Hamaker coefficient is safely reached after about $10$ layers, the swiftness of the transition to the continuum approximation being dictated by the thickness and the dielectric properties of the layers $A$ and $B$. Also the magnitude of the Hamaker coefficient, for fixed thickness of the periodic layer, $a + b = 8 nm$ in the case examined, depends on the volume fraction of both layers. It is biggest when the lipid layer volume fraction is highest. This is indeed as one would expect since for high water volume fractions the dielectric inhomogeneity and thus van der Waals interactions are non-existent. It is only at much higher values of $N$ that the effect of retardation will set in, leading to a levelling off of the Hamaker coefficient at about half of its $n = 0$ value. Since our focus here is the granularity or stratification of the medium, we do not delve any further on the retardation effects.

If the Hamaker coefficient were evaluated in the model based on an isotropic mixture of $A$ and $B$ the Hamaker coefficient would be about $20 \%$ smaller than in the anisotropic continuum model.

Comparing  the exact dependence of the van der Waals interactions across a periodic multilayer on the number of layers with a continuum approximation based on the van der Waals interactions across uni-axial anisotropic composite, we have been able to assess the validity of the continuum approximation and show that it indeed reduces to the properly defined limit after about $10$ layers. Again, this limit has to acknowledge the uniaxial continuum structure of the multilayer.


\begin{thebibliography}{9}

\bibitem{nagle1} Nagle JF, Tristram-Nagle S \textsl{BBA-Rev Biomembranes}  \textbf{1469} (2000 ) 159-195

\bibitem{nagle2} Nagle JF, Katsaras J \textsl{Phys Rev E} \textbf{59}  (1999) 7018-7024 

\bibitem{decher1} Decher G, Eckle M, Schmitt J, Struth B
\textsl{Curr Op Colloid Interface Sci,}  \textbf{3 } (1998) 32-39 

\bibitem{sukhishvili} Sukhishvili SA, Granick S
\textsl{Macromol} \textbf{35}  (2002) 301-310 

\bibitem{decher2} Decher G \textsl{Science} 
 \textbf{277} (1997) 1232-1237 

\bibitem{rudi1}  R. Podgornik, P. L. Hansen, and V. A. Parsegian
\textsl{J. Chem. Phys.} \textbf{119}  (2003) 1070-1077  

\bibitem{optics} M. Born and E. Wolf, {\sl Principles of optics} (The Macmillan Company, New York) 1964, Ch. 1.6.5.

\bibitem{Abeles} F. Abel\` es, \textsl{Ann. de Physique} \textbf{5} (1950) 777. 

\bibitem{adrian1}  B.W. Ninham and V.A. Parsegian \textsl{J Chem Phys} 
\textbf{53} (1970) 3398-3402

\bibitem{ninham} J. Mahanty and B.W. Ninham, {\sl Dispersion forces} (Academic Press, New York) 1976.

\bibitem{sarlah} \v Sarlah A. and \v Zumer S., {\sl Phys. Rev. E} {\bf 64} (2001) 051606-1.

\end{thebibliography}
\end{document}